\documentclass[osajnl,twocolumn,showpacs,superscriptaddress,11pt]{revtex4-1} 
\pdfoutput=1
\usepackage{amsmath,amssymb,graphicx}
\newcommand{\bxi}{\boldsymbol{\xi}}
\newcommand{\brho}{\boldsymbol{\rho}}
\newcommand{\brhoc}{\boldsymbol{\rho_c}}
\newcommand{\brhod}{\boldsymbol{\rho_d}}

\newcommand{\bkp}{\boldsymbol{\kappa_\perp}}
\newcommand{\bkpc}{\boldsymbol{\kappa_{\perp c}}}
\newcommand{\bkpd}{\boldsymbol{\kappa_{\perp d}}}

\begin{document}

\title{Field of view advantage of conjugate adaptive optics in microscopy applications}

\author{Jerome Mertz}\email{Corresponding author: jmertz@bu.edu}
\affiliation{Dept. of Biomedical Engineering, Boston University, 44 Cummington Mall, Boston, MA 02215, USA}

\author{Hari Paudel}
\affiliation{Dept. of Electrical Engineering, Boston University, 8 Saint Mary's St., Boston, MA 02215, USA}

\author{Thomas G. Bifano}
\affiliation{Photonics Center, Boston University, 8 Saint Mary's St., Boston, MA 02215, USA}

\begin{abstract}
The imaging performance of an optical microscope can be degraded by sample-induced aberrations. A general strategy to undo the effect of these aberrations is to apply wavefront correction with a deformable mirror (DM). In most cases, the DM is placed conjugate to the microscope pupil, called pupil adaptive optics (AO). When the aberrations are spatially variant, an alternative configuration involves placing the DM conjugate to the main source of aberrations, called conjugate AO. We provide theoretical and experimental comparison of both configurations for the simplified case where spatially variant aberrations are produced by a well defined phase screen. We pay particular attention to the resulting correction field of view (FOV). Conjugate AO is found to provide a significant FOV advantage. While this result is well known in the astronomy community, our goal here is to recast it specifically for the optical microscopy community.   
\end{abstract}


\maketitle 

\section{Introduction}

Objects become blurred when they are imaged through scattering media. This has been a longstanding source of frustration in optical microscopy, particularly in biomedical imaging applications where objects of interest are routinely embedded within scattering media or behind aberrating surfaces. A well-known strategy to counteract aberrations makes use of adaptive optics (AO), as borrowed from astronomical imaging \cite{tyson, roddier}. The idea of adaptive optics is to insert an element in the imaging optics, typically a deformable mirror (DM), that imparts inverse aberrations to the imaged light, thus compensating for the aberrations induced by the sample or by the microscope system itself. In most cases, the DM is inserted in the pupil plane (or conjugate plane thereof) of the microscope optics. Such a DM placement is appropriate when the aberrations to be compensated are spatially invariant, as in the case when they are produced by an index of refraction mismatch at a flat interface. But another reason for placing the DM in the pupil plane seems more historical in nature, and is largely a carryover from astronomical AO. However, one should bear in mind that the requirements for astronomical compared to microscopy AO are very different. In astronomical AO one is usually interested in only very small (angular) fields of view and it is important to bring all the DM actuators to bear on single, localized objects at a time. In microscopy AO, particularly involving widefield (i.e. non-scanning) configurations, the opposite is usually true and it is desirable perform AO over as large a field of view (FOV) as possible. In the more general case when sample-induced aberrations are spatially variant as opposed to invariant, a placement of the DM in the pupil plane turns out to be a very poor choice as it imposes a severe limitation on FOV \cite{fried_seeing}. A better choice is to place the DM in a plane conjugate to the plane where the aberrations are the most dominant, and hence deleterious. This is called conjugate AO. The purpose of this manuscript is to highlight the differences between conjugate and pupil AO, particularly where FOV is concerned, both theoretically and experimentally.

To begin, we emphasize that what will be said here is not new. It is well known \cite{beckers,welsh,ragazzoni,sarazin}  in the astronomical imaging community that the FOV (or ``seeing") can be extended with the use of conjugate AO. In this case, it is assumed that  the most important aberrations are produced by turbulence from a well defined layer in the atmosphere, and the DM is placed conjugate to this layer. This strategy can be generalized to multi-conjugate adaptive optics (or MCAO) where multiple DMs are conjugated to multiple atmospheric layers. 

The principle of MCAO, though well understood in the astronomy community, seems to be less appreciated in the microscopy community \cite{booth_review}. A few reports have discussed various benefits of MCAO in the context of microscopy \cite{sedat, booth}, though these have relied on numerical simulation only. MCAO has also been used in retinal imaging applications \cite{retina} and in benchtop experiments designed to simulate astronomical imaging \cite{dainty}. Our goal here is to build on these results by providing a theoretical framework specifically tailored to the microscopy community. We limit our considerations to the simplified case where spatially variant aberrations are assumed to arise from a single layer only. While such a case may seem overly idealized, it serves to highlight the salient features of conjugate AO regarding FOV, which is our goal here. It also becomes relevant, for example, in sub-surface imaging applications where the dominant aberrations arise from irregularities at the sample surface, as is common in practice.

Our manuscript is organized as follows. We begin by describing the effect on imaging of an aberrating layer an arbitrary distance from the focal plane. We then consider the effects of compensating these aberrations by placing a DM first in the pupil plane and then in the plane conjugate to the aberrating layer. We concentrate our discussion on the implications for FOV. Finally, in the second half of this manuscript, we support our theoretical results with proof of principle experiments involving image-based AO with a calibrated object and a biological sample, for demonstration purposes. Our goal is to lay the groundwork for future bona-fide microscopy applications.     

\section{Effect of a single aberrating layer}
We consider a telecentric microscope system, which, for simplicity, we take to have unit magnification (see Fig.~1). The complex object field located at the focal plane is given by $E_0(\brho)$, where $\brho$ is a 2D lateral coordinate. This field is assumed to be quasi-monochromatic of average wavelength $\lambda$. Field propagation through the microscope is taken to be through free space, except for the presence of a thin aberrating layer located a distance $z$ from the object, modeled as a thin phase screen of transmission $t(\brho)=e^{i\phi(\brho)}$, where $\phi(\brho)$ is a local (real) phase shift.
Throughout this discussion we will adopt the paraxial, or Fresnel, approximation, meaning we will consider only propagation angles that are small. Implicitly, this means we assume our phase screen is forward scattering only, meaning the lateral extent of its phase features is typically larger than $\lambda$ (more on this later).

\begin{figure}[h!]
\centerline{\includegraphics[width=.9\columnwidth]{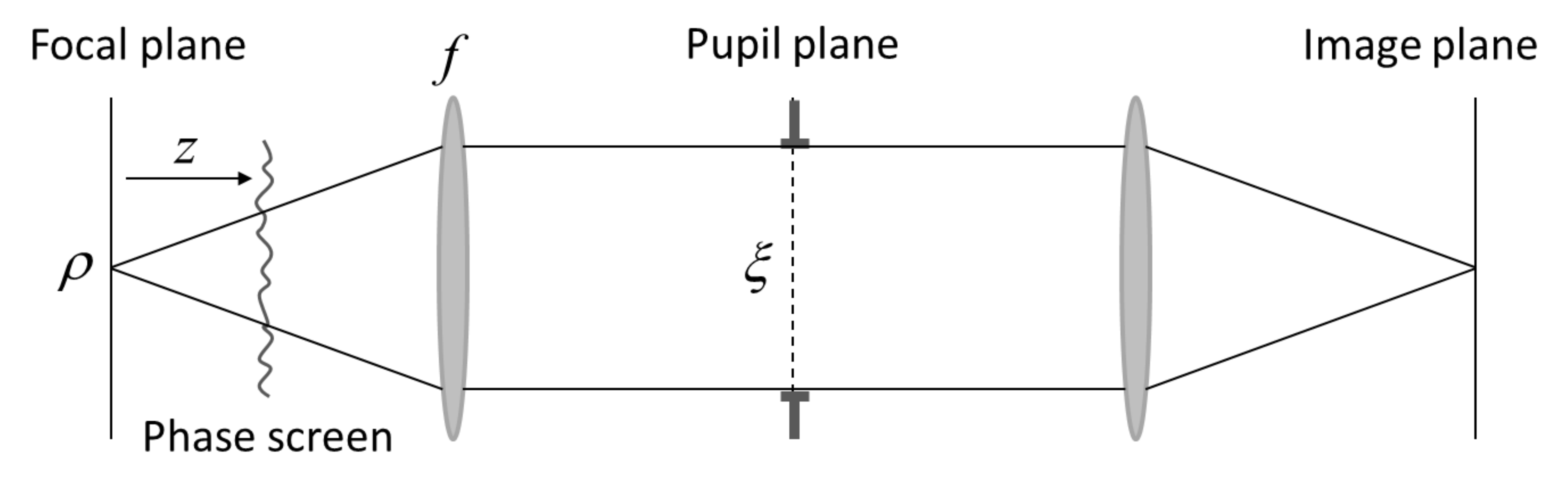}}
\caption{Basic microscope layout.}
\end{figure}

As far as our imaging device is concerned, the object field $E_0(\brho)$ propagating through the phase screen $t(\brho)$ is equivalent to an effective, albeit scrambled, field $E(\brho)$ propagating through free space. We can derive $E(\brho)$ by propagating $E_0(\brho)$ to the phase screen, multiplying it by $t(\brho)$, and propagating it back to the focal plane. Using standard Fresnel propagation integrals, we find    

\begin{equation}
E(\brho)=\iint{e^{i2\pi\bkp\cdot(\brho-\brho')}
E_0(\brho+z\lambda\bkp)t(\brho')d^2\brho' d^2\bkp},
\end{equation}

\noindent where $\bkp$ is a 2D transverse spatial frequency and $\lambda\bkp$ may be interpreted as a propagation angle (we have adopted similar notation as in \cite{mertz}). Again, in keeping with the paraxial approximation, the integral over $\bkp$ is assumed to span a range such that $\lambda\bkp\ll 1$. In reality, the range of $\bkp$ will become even more restricted by our microscope pupil, but we do not consider this yet. Equation~1 is taken to be valid independent of our imaging device and will be the starting point of our discussion.

Ultimately, we will use a camera to form an image, and thus we are interested in recording intensities rather than fields. With this in mind, we evaluate the mutual intensity of $E(\brho)$, defined as $J(\brho_c,\brho_d)=\langle E(\brho_c+\tfrac{1}{2}\brho_d)E^*(\brho_c-\tfrac{1}{2}\brho_d)\rangle$, where $\langle ...\rangle$ corresponds to a time average. A tedious calculation yields

\begin{equation}
\begin{split}
J(\brhoc,\brhod)&=\iiiint{e^{i2\pi\bkpd\cdot(\brhoc-\brhoc')}e^{i2\pi\bkpc\cdot(\brhod-\brhod')}}\\
&J_0(\brhoc-z\lambda\bkpc,\brhod-z\lambda\bkpd)\Gamma_0(\brhoc',\brhod')\\
&d^2\brhoc' d^2\brhod' d^2\bkpc d^2\bkpd,
\end{split}
\end{equation}
where we have introduced the function  $\Gamma_0(\brho_c,\brho_d)= t(\brho_c+\tfrac{1}{2}\brho_d)t^*(\brho_c-\tfrac{1}{2}\brho_d)$.

So far, we have made no assumptions regarding the object field $E_0(\brho)$. We now make the assumption that it is spatially incoherent, as is the case, for example, when imaging fluorescence. That is, we formally write $J_0(\brhoc,\brhod)\approx \lambda^2 I_0(\brhoc) \delta^2(\brhod)$, where the prefactor $\lambda^2$ is introduced for dimensional consistency, but also because the coherence area of radiating spatially incoherent fields is roughly $\lambda^2$.  Equation~2 then simplifies to
 
\begin{equation}
\begin{split}
J(\brhoc,\brhod)&=\tfrac{1}{z^2}\iiint{e^{i2\pi\brhod\cdot(\brhoc-\brhoc')/z\lambda}e^{i2\pi\bkpc\cdot(\brhod-\brhod')}}\\
&I_0(\brhoc-z\lambda\bkpc)\Gamma_0(\brhoc',\brhod')d^2\brhoc' d^2\brhod' d^2\bkpc.
\end{split}
\end{equation} 

In other words, even though the object field $E_0(\brho)$ is taken to be spatially incoherent, the apparent field $E(\brho)$ may develop spatial coherences owing to the presence of the phase screen, a result that stems, in part, from the Van Cittert-Zernike theorem (e.g. see \cite{mertz}). Let us examine these spatial coherences more closely. To do this, we make some assumptions regarding the phase screen.

Up to this point, our calculations have been similar to those encountered in astronomical imaging through a turbulent atmosphere \cite{roddier}. In the latter case, the phase fluctuations imparted by $t(\brho)$ are assumed to be random in time and long image exposures are taken such that $\Gamma_0(\brho_c,\brho_d)$ can be reduced to its wide-sense stationary representation $\bar\Gamma_0(\rho_d)$, which is independent of $\brho_c$. We cannot perform such time averaging here because, in contrast to atmospheric turbulence, our phase screen is assumed to be static. Nevertheless, $\Gamma_0(\brho_c,\brho_d)$ is not isolated in Eq.~3, but rather occurs inside an integral. We make an approximation of ergodicity by writing

\begin{equation}
\int{e^{-i2\pi\bkp\cdot\brhoc}\Gamma_0(\brhoc,\brhod)d^2\brhoc}\approx
\delta^2(\bkp) \bar\Gamma_0(\rho_d).  
\end{equation}     

\noindent In effect, we assume that the phase screen is situated far enough from the focal plane that light arising from any object point traverses a phase-screen area large enough to encompass many uncorrelated, statistically homogeneous phase features. With this approximation, $E(\brho)$ also becomes spatially incoherent, and Eq.~3 reduces to

\begin{equation}
I(\brho)\approx\iint{e^{-i2\pi\bkp\cdot\brho'}
I_0(\brho+z\lambda\bkp)\bar\Gamma_0(\rho')d^2\brho' d^2\bkp}.
\end{equation} 

We recall that $I(\brho)$ is the apparent object intensity at the focal plane resulting from the propagation of the actual object intensity $I_0(\brho)$ through the phase screen.  This equation bears resemblance to and is essentially the intensity equivalent of Eq.~1, valid for spatially incoherent object fields. We note that we still have not considered the imaging device in our calculations. However, when we do, because the apparent object field remains spatially incoherent, we need only invoke the intensity point spread function (PSF), as opposed to the amplitude PSF, to evaluate the resultant image. 

To make further progress, we must make some assumptions regarding $\bar\Gamma_0(\rho)$. Conventionally \cite{goodman}, this is done by writing $\bar\Gamma_0(\rho_d)=\langle$exp$(i\phi(\brhoc+\tfrac{1}{2}\brhod)-i\phi(\brhoc-\tfrac{1}{2}\brhod))\rangle$, where $\langle...\rangle$ now refers to an average over $\brhoc$. Assuming $\phi(\brho)$ are Gaussian random processes of zero mean, we find $\bar\Gamma_0(\rho_d)=$exp$(-\tfrac{1}{2}D(\rho_d))$ where $D(\rho_d)=\langle(\phi(\brhoc+\tfrac{1}{2}\brhod)-\phi(\brhoc-\tfrac{1}{2}\brhod))^2\rangle$ is the structure function of the phase variations. Taking these variations to be statistically homogeneous, and characterized by a variance ${\sigma_\phi}^2$ and normalized spatial autocorrelation function $\gamma_0(\rho_d)$, we obtain   
 
\begin{equation}
\bar\Gamma_0(\rho_d)=e^{-{\sigma_\phi}^2 (1-\gamma_0(\rho_d))},
\end{equation}

\noindent or, equivalently,

\begin{equation}
\bar\Gamma_0(\rho_d)=e^{-{\sigma_\phi}^2 }+e^{-{\sigma_\phi}^2} (e^{{\sigma_\phi}^2 \gamma_0(\rho_d)}-1).
\end{equation}

The advantage of Eq.~7 is that it identifies the effects of the phase screen on ballistic (first term) and scattered (second term) light propagation. We adopt the model here that the phase variations are correlated in a Gaussian manner over a characteristic length scale $l_\phi$, such that $\gamma_0(\rho_d)=e^{-\rho_d^2/l_\phi^2}$. Plots of $\gamma_0(\rho_d)$ are illustrated in Fig.~2 for various values of ${\sigma_\phi}^2$. 

\begin{figure}[h!]
\centerline{\includegraphics[width=.9\columnwidth]{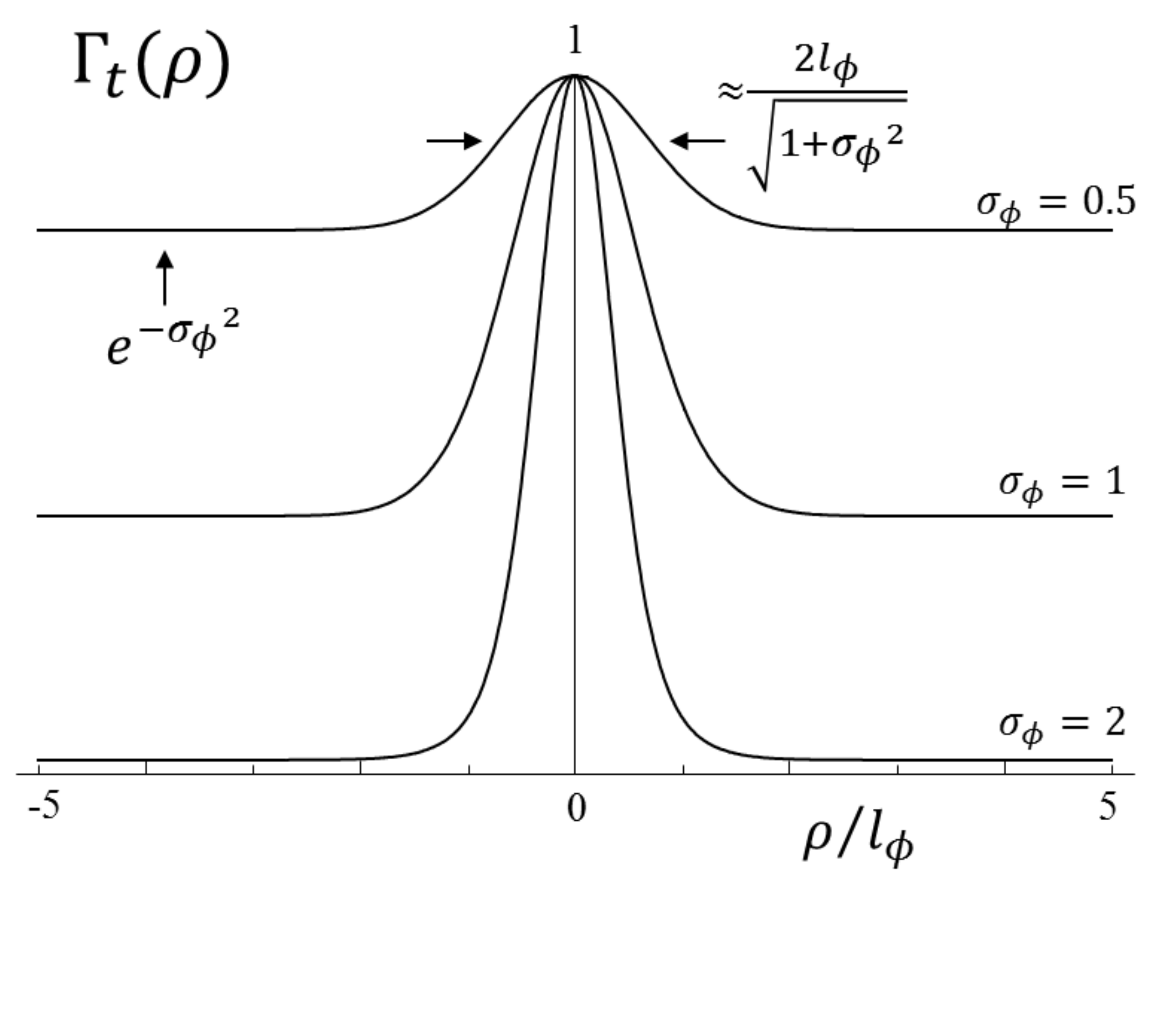}}
\caption{Plots of $\Gamma_0(\rho)$ for different values of ${\sigma_\phi}^2$.}
\end{figure}

To gain further insight into the problem, we take Eq.~7 one step further by recasting it yet again in the form

\begin{equation}
\bar\Gamma_0(\rho_d)\approx e^{-{\sigma_\phi}^2 }+(1-e^{-{\sigma_\phi}^2}) \gamma_0(\rho_d \sqrt{1+{\sigma_\phi}^2}).
\end{equation}  

Equation~8 is found to be an excellent approximation to Eq.~7 for values of ${\sigma_\phi}^2$ much smaller and much larger than one, and only deviates slightly from Eq,~7 when ${\sigma_\phi}^2$ is close to one. We thus adopt it as a general expression for $\Gamma_0(\rho_d)$ since it provides much clearer insight into its physical meaning.

As an example, let us consider the object to be a point source located at the origin of the focal plane. That is, we write $I_0(\brho)=\lambda^2 I_0 \delta^2(\brho)$. Plugging this into Eq.~ 5 , we find the effective object intensity to be

\begin{equation}
I(\brho)=\tfrac{1}{z^2}I_0\int{e^{-i2\pi\brho\cdot\brho'/z\lambda}\bar\Gamma_0(\rho')d^2\brho'}.
\end{equation} 

Propagating this intensity through our imaging device and making use of Eq.~9, we derive an effective PSF given by
 
\begin{equation}
\textrm{PSF}(\brho)=e^{-{\sigma_\phi}^2 }\textrm{PSF}_0(\brho)+(1-e^{-{\sigma_\phi}^2})\frac{1}{\pi\zeta^2}
e^{-\rho^2/\zeta^2},
\end{equation}

\noindent where PSF$_0(\brho)$ is the initial device PSF absent the phase screen, and we introduce the length scale $\zeta=z \lambda \sqrt{1+{\sigma_\phi}^2}/\pi l_\phi$. A plot of this effective PSF is illustrated in Fig.~3.  We observe that it features two components. The first is an attenuated version of the initial PSF$_0$, leading to a sharp, diffraction-limited peak that is attenuated because of loss of ballistic light due to scattering. The second is a broad pedestal resulting from scattering due to the phase screen. An increase in $\zeta$ caused, for example, by an increase in $z$ leads to an increase in the pedestal width and a concomitant decrease in its height. The proportion of the effective PSF that remains ballistic is $e^{-{\sigma_\phi}^2 }$, while the rest is scattered. Clearly, as ${\sigma_\phi}^2$ increases the effective PSF becomes progressively more diffuse. 

\begin{figure}[h!]
\centerline{\includegraphics[width=.9\columnwidth]{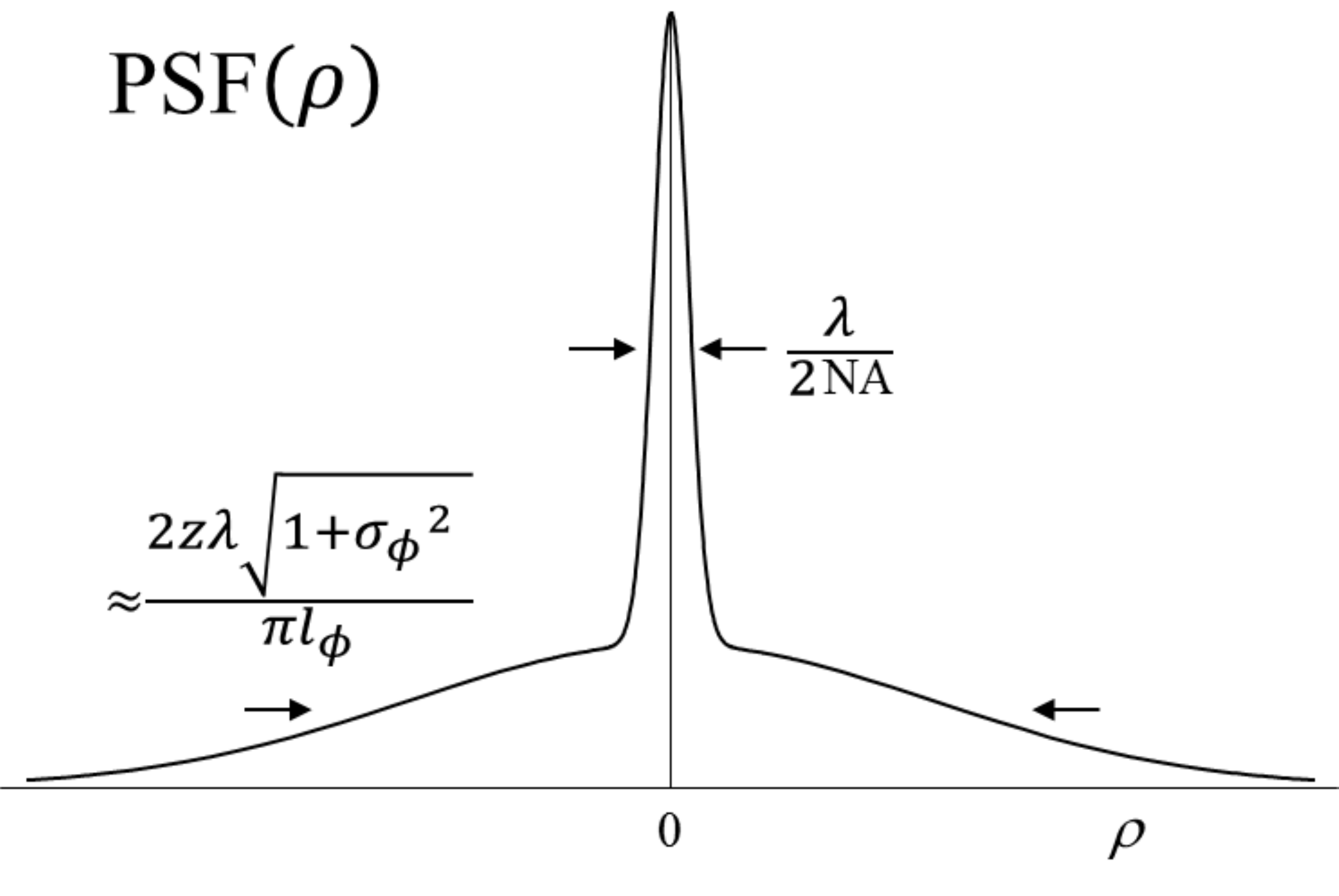}}
\caption{Example of an effective PSF$(\rho)$ after degradation by a phase screen. The initial diffraction-limited PSF (PSF$_0$) is reduced by a factor $e^{-{\sigma_\phi}^2}$ and rides on a broader  blurred background of power $1-e^{-{\sigma_\phi}^2}$. }
\end{figure}

Some comments on the validity of Eq.~10. To begin, we recall that we have limited ourselves to small propagation angles. $\zeta$ must therefore be smaller than $z$, meaning we implicitly assume $\lambda \sqrt{1+{\sigma_\phi}^2}/l_\phi \ll 1$. Moreover, in assuming that the light from the point object samples many phase-screen correlation areas, we implicitly restrict the microscope NA to values much larger than $l_\phi/z$. In turn, this means that $\zeta$ is implicitly assumed to be larger than the diffraction limited spot size $\lambda/2$NA. Bearing these assumptions in mind, we may evaluate the Strehl ratio of our effective PSF , defined by $S=$PSF$(0)/$PSF$_0(0)\approx\lambda^2$PSF$(0)/\pi$NA$^2$. For phase variances not so large that the Strehl ratio is defined primarily by ballistic light we obtain $S=e^{-{\sigma_\phi}^2 }$. This result is the same as obtained for astronomical imaging through a turbulent atmospheric layer \cite{tyson}.  

How do we recover from the degradation in PSF$(\brho)$ caused by the phase screen? We turn now to two possible strategies involving different AO configurations.    

\section{Pupil AO} 

The most common AO configuration involves placing a DM in the pupil plane of the imaging device. Wavefront correction is then based either on knowledge gained from a direct measurement of the aberrated wavefront at this plane (wavefront-sensing-based AO), or by iterative trial and error to optimize a user-defined image metric (image-based AO). In either case, wavefront correction is designed to undo the effects of aberrations for a particular spot in the object, typically the location of a ``guide star" \cite{tyson}, with the hope that the range of the correction about this spot is large enough to encompass neighboring objects of interest. In astronomy parlance, this range is called ``seeing" or the ``isoplanatic patch".  Here, we call it the corrected FOV of our microscope. Our goal in this section is to derive an expression for this FOV.   

To begin, we assume that the object spot whose wavefront we would like to correct is located at the origin. This could be the location of a fluorescent beacon (serving as a guide star), or of an object point we have arbitrarily selected for image optimization. To evaluate the aberrated wavefront produced by this spot, we write the object field at this spot as $E_0(\brho)=\lambda^2 E_0 \delta^2(\brho)$. Inserting this into Eq.~1, and performing a scaled Fourier transform to calculate the resultant field at the pupil plane, we find  

\begin{equation}
E_p(\bxi)=\frac{\lambda}{f}E_0\iint{e^{-i2\pi\bkp\cdot(\brho+\tfrac{z}{f}\bxi)}
e^{i\pi z\lambda\kappa_\perp^2}t(\brho)d^2\brho d^2\bkp},
\end{equation}

\noindent where $\bxi$ is a 2D transverse coordinate in the pupil plane and $f$ is the focal length of the microscope objective (see Fig.~1). We observe that in the absence of a phase screen ($t(\brho)=1$), $E_p(\bxi)$ becomes a plane wave of constant amplitude $\lambda E_0/f$ that is uniformly spread across the pupil plane, as expected. In contrast, when the phase screen is present $E_p(\bxi)$ becomes structured both in amplitude and in phase. To correct for this effect of the phase screen, we can act on $E_p(\bxi)$ with a wavefront corrector. Ideally, $E_p(\bxi)$ should be made into a plane wave again, meaning  we should flatten it both in amplitude and phase. However, wavefront correctors such as DMs are never ideal and generally act on phase only. The best we can do is to flatten the phase of $E_p(\bxi)$ and leave its amplitude unchanged. Nevertheless, as we will argue below, the consequences of applying phase-only wavefront correction are not all that severe.  

In effect, the DM is itself a phase screen whose transmission function can be written as $t_{\textrm{dm}}(\bxi)$. To correct for the phase variations in $E_p(\bxi)$, the DM should impart its own phase variations that are precisely the negative of those of  $E_p(\bxi)$. That is, the DM should effectively phase conjugate $E_p(\bxi)$. This occurs when

\begin{equation}
t_{\textrm{dm}}(\bxi)=\frac{E^*_p(\bxi)}{\lvert E_p(\bxi) \rvert}.
\end{equation} 

At this point, the math becomes difficult and we need to introduce a simplification to proceed. Specifically, we replace the denominator in Eq.~12 with the (constant) field amplitude that would be present without the phase screen, and write

\begin{equation}
t_{\textrm{dm}}(\bxi)=\iint{e^{i2\pi\bkp\cdot(\brho+\tfrac{z}{f}\bxi)}
e^{-i\pi z\lambda\kappa_\perp^2}t^*(\brho)d^2\brho d^2\bkp}.
\end{equation}
 
This simplification has no effect on the average intensity at the pupil plane, and we may justify it on the grounds that, with the aid of the approximation in Eq.~4, we maintain $\lvert t_{\textrm{dm}}(\bxi) \rvert \approx1$, in agreement with the phase-only nature of our DM.  Nevertheless, this simplification does somewhat modify the field statistics at the pupil plane, the ramifications of which will be discussed below.

We recall that $t_{\textrm{dm}}(\bxi)$ here is the shape applied to the DM that performs optimal AO correction exactly at the origin of the focal plane.  We can now evaluate the spatial range, or FOV, over which this correction remains effective. To do this, we begin now with an arbitrary (albeit spatially incoherent) object field. The effective field at the object plane that takes into account the aberrations imparted  by phase screen is given by $E_(\brho)$ in Eq.~1. To take into account the additional, hopefully corrective, effect of the DM, we propagate this field to the pupil (a scaled Fourier transform), multiply it by $t_{\textrm{dm}}(\bxi)$, and then propagate it back to the focal plane (a scaled inverse Fourier transform). Using the same math and associated approximations as in the previous section, and keeping only dominant phase correlations, we arrive at

\begin{equation}
I_{\textrm{ao}}(\brho)\approx I_0(\brho) \bar\Gamma^2_t(\brho).
\end{equation}  

Again, this result is subject to the same conditions of validity as before. Moreover, we have assumed that the DM spans the entire pupil (not too small), and its actuators are sufficiently dense to accurately represent $t_{\textrm{dm}}(\bxi)$ (more on this below). 

Equation 14 warrants scrutiny. $I_0(\brho)$ is the actual object intensity; $I_{\textrm{ao}}(\brho)$ is the effective object intensity as observed through the phase screen and corrected by the pupil DM. The window of AO correction is thus characterized by $\bar\Gamma^2_t(\brho)$. From Eq.~6, we note that this window is equivalent to $\bar\Gamma_0(\brho)$, but with a phase variance ${\sigma_\phi}^2$ that is effectively doubled. We can thus recast it using the approximation provided by Eq.~8. However, it is at this point that we introduce a small correction to our results. Namely, we write
    
\begin{equation}
\bar\Gamma^2_t(\rho_d)\approx e^{-2{\sigma_\phi}^2 }+(1-e^{-2{\sigma_\phi}^2}) \frac{\pi}{4}\gamma_0(\rho_d \sqrt{1+2{\sigma_\phi}^2}).
\end{equation}  

The reader can observe that we have introduced an extra factor of $\pi/4$ in the second term.  The justification for this term is as follows. We recall that the reason for separating $\bar\Gamma_0(\rho_d)$ into two terms was to better identify the  ballistic and scattering propagation components. The second term arises from light that is scattered by the phase screen. For light that originates from a point at the focal plane, the scattered component of this light then impinges the pupil plane with spatially varying amplitude and phase. Indeed this scattered component takes on the characteristics of fully developed speckle \cite{goodman} (fully developed because we have removed from it all ballistic contribution). Upon correcting the wavefront of this speckle field, the best the DM can do is flatten its phase. The amplitude distribution of the speckle field remains unchanged, meaning it continues to obey Rayleigh statistics. In the typical case that the microscope pupil is physically smaller than the field distribution at the pupil plane (i.e. there is some energy loss), then the best the DM can do is yield a corrected image intensity from this scattered light component that is $\pi/4$ reduced compared to the same component with no phase screen \cite{mosk}. We need to include this correction factor in  Eq.~15 because our derivation made use of Eq.~13 for the optimal DM profile rather than the more correct Eq.~12.  

As an aside, we note that the spatial extent of the speckle grains at the pupil plane is inversely related to the spatial extent of the aberrated intensity pattern at the object plane, and is thus given by, roughly, $\lambda f/2 \zeta$ (akin to the Fried parameter in astronomical imaging \cite{fried_r0}). The DM actuators must be smaller than this to avoid undersampling of $t_p(\bxi)$. 

\begin{figure}[h!]
\centerline{\includegraphics[width=.9\columnwidth]{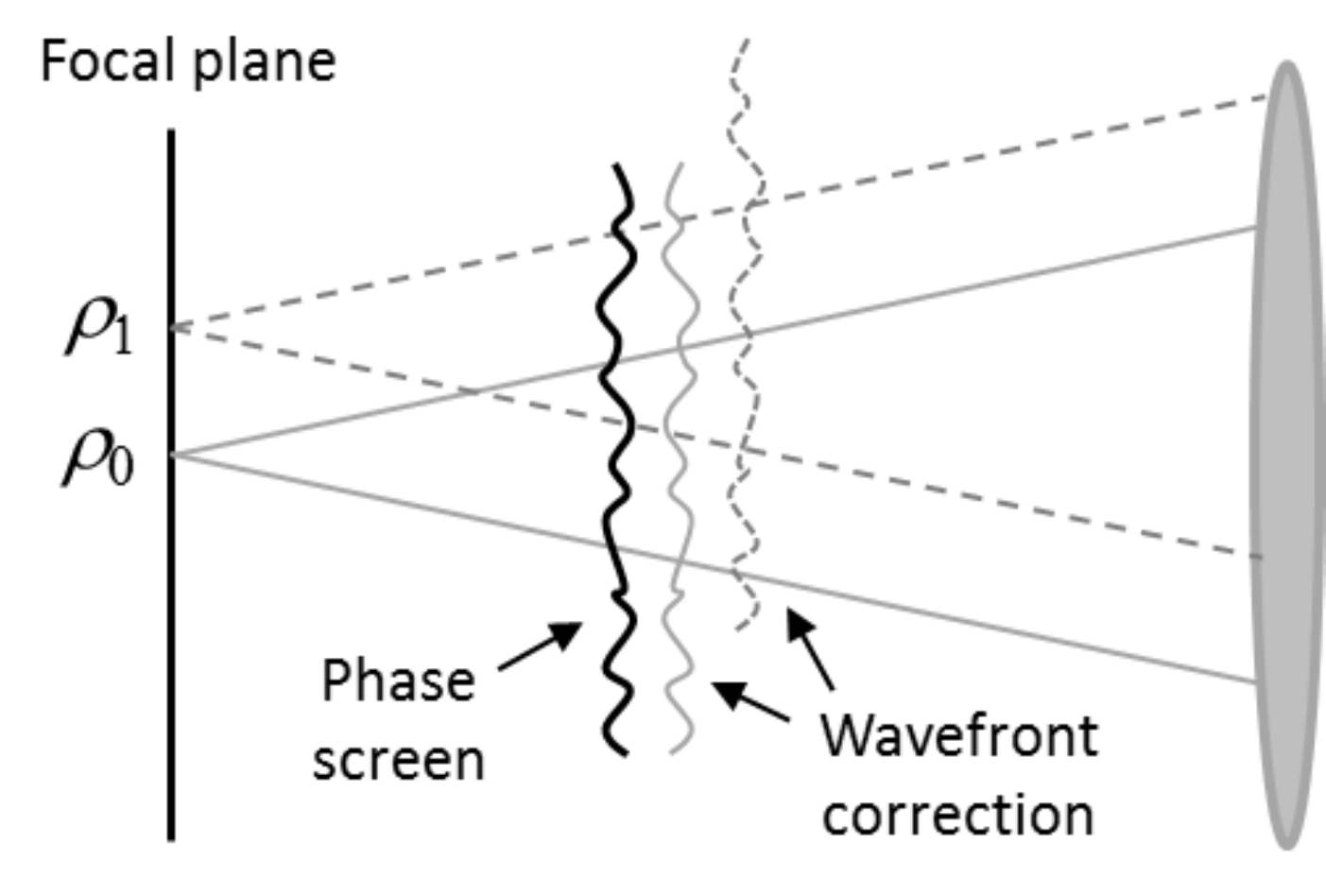}}
\caption{Interpretation of what defines FOV for pupil AO.  As an object point $\brho_1$ becomes displaced from the guide-star point $\brho_0$,  its wavefront correction becomes uncorrelated with the fixed phase screen aberrations, and AO fails. }
\end{figure}

To summarize, $\bar\Gamma^2_t(\rho_d)$ represents the window over which pupil-based AO is effective. For phase variations of  standard deviation $\sigma_\phi$ one radian or larger, this window is predominantly defined by the second term and yields a FOV of diameter $a_0\approx 2l_\phi/\sqrt{1+2{\sigma_\phi}^2}$. This FOV becomes narrower with increasing phase variations, approaching the limit of $\sqrt{2}l_\phi/\sigma_\phi$. In other words, in the regime where the phase variations are large (i.e. the regime where AO is potentially most useful!) the FOV depends not on the characteristic scale $l_\phi$ of the aberrating features, but rather on their (inverse) characteristic slope. This result is similar to a general result for imaging through phase screens discussed in \cite{goodman}. Moreover, the FOV is independent of the location $z$ of the phase screen. These observations can readily be understood from Fig.~4. We recall that any wavefront correction provided by a pupil-plane DM must be spatially invariant in the sense that it must be imparted equally to all object points. The  wavefront correction can be thought of, therefore, as figuratively tracking each object point. However, for object points displaced from the guide star the wavefront correction becomes rapidly uncorrelated with the aberrations it is intended to correct. In this case, the correction actually produces worse imaging than if there were no DM at all, because it leads to the presence of two uncorrelated phase screens in the imaging optics rather than one (hence the effective doubling of ${\sigma_\phi}^2$). Because the microscope system is telecentric, this FOV is independent of $z$. 

We close this section with a reminder that our derivation of $a_0$ presumed that pupil AO was asked to optimize the image at a single point only, namely at the origin. One might wonder what would happen if it were asked to optimize over a larger region, say of size $A$. This problem is tantamount to optimizing multiple guide stars simultaneously. Such simultaneous optimization is possible; however, it is known to lead to reduced contrast enhancement at each guide star by a factor of $N$, the number of guide stars \cite{mosk}. That is, while pupil AO can, in principle , correct over a range $A$ larger than $a_0$, the quality of this correction, as measured by contrast enhancement, is expected to rapidly degrade as $(a_0/A)^2$.        

\section{Conjugate AO} 
In the previous section we characterized a problem (often debilitating) of pupil-plane AO that it cannot provide extended FOVs. This disadvantage was interpreted as arising from the property that the pupil-plane AO wavefront correction effectively tracks the different points in the object plane while the aberrating phase screen remains fixed. A solution to this problem is clear. The wavefront correction should instead be locked to the phase screen rather than to the object points. This can be achieved by placing the DM conjugate to the phase screen itself, called conjugate AO.

\begin{figure}[h!]
\centerline{\includegraphics[width=.9\columnwidth]{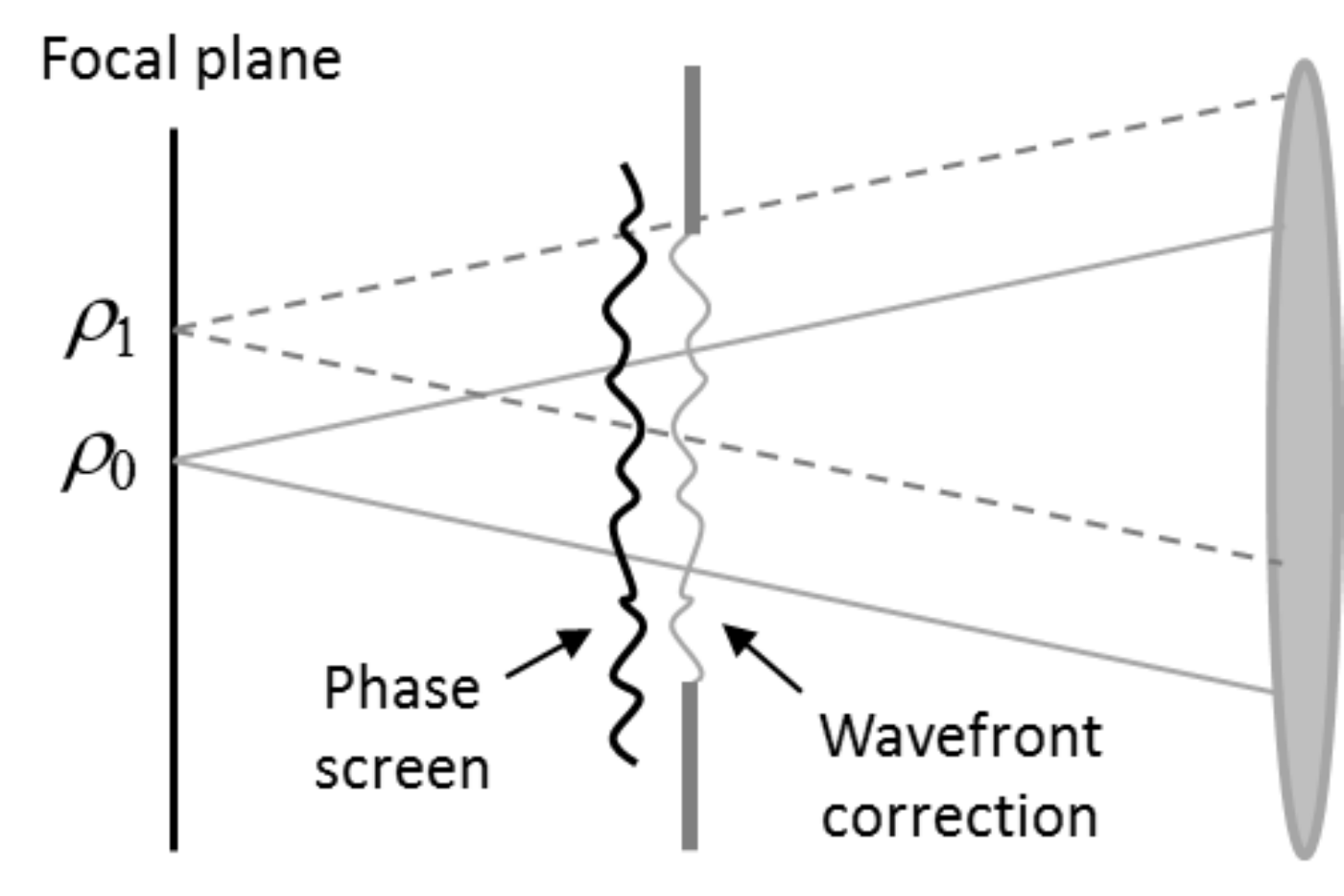}}
\caption{Interpretation of what defines FOV for conjugate AO.  The fixed wavefront correction cancels the aberrations caused by the fixed phase screen. The FOV is then limited by size of the projected DM itself, albeit with blurred edges dependent on distance $z$ of the phase screen and the microscope NA.}
\end{figure}

What is the FOV associated with conjugate AO? Based on the above interpretation, it is clear that the FOV must be the size of the DM projected onto the phase screen. For example, if the DM is conjugate to the phase screen with unit magnification, then the FOV is the same size as the DM itself. Again, since our microscope system is telecentric, this FOV is independent of the $z$ location of the phase screen, provided the DM is maintained conjugate. However, one must be careful with this last statement since there may be edge effects that are dependent on $z$. Such effects are best appreciated with a simple ray optic picture as shown in Fig.~5. The microscope pupil defines a characteristic maximum cone size (i.e. NA) of light that can be collected from any object point. If this cone size is large enough that it spans the projected DM for all object points within the FOV, then conjugate AO corrects equally well throughout this FOV.  On the other hand, if the cone size is smaller (as shown), then the DM causes vignetting. This vignetting begins to occur at distances roughly $z$NA from the FOV edges (note:  the DM here is assumed to be obstructing beyond its edges). Nevertheless, despite this potential issue of edge effects, it is clear (and well known \cite{beckers}) that the FOV provided by conjugate AO can be significantly larger than that provided by pupil AO.  

\section{Experimental demonstration}

The FOV advantage of conjugate AO can be demonstrated experimentally. In most cases, AO is applied in laser scanning microscope configurations \cite{wilson, girkin,denk,miller,beaurepaire,vellekoop,boppart,kubby,cui,ji,betzig}. We apply it here instead in a bright-field configuration. Our setup is illustrated in Fig.~6 
and is basically a $1.33\times$ magnification microscope modified to accommodate a pupil or conjugate DM (Boston Micromachine Corp. MultiDM, 
140 actuators in a square $12\times12$ array without the corner actuators, $400 \mu$m actuator pitch). A thin transmission object is trans-illuminated by a LED (660~nm). A thick diffuser is inserted in the LED path, just  before the object, to ensure object spatial incoherence. The NA of the microscope is about 0.04, as defined  by a $\approx$8~mm diameter iris pupil. We note that there is a factor of about two demagnification in the imaging optics from this pupil to the pupil DM to properly match their respective sizes. A mirror blank is inserted in the place of the conjugate DM when the system is in a pupil DM configuration, and vice versa. The camera is a Thorlabs DCC1545M CMOS (pixel size $5.2\mu$m square).

To introduce spatially variant aberrations in our microscope, we inserted a phase screen a distance $z$ (here 35~mm) from the object. The phase screen consisted of a blank microscope slide onto which was spray-painted a thin layer of clear acrylic. The profile of this phase screen was measured by a Zygo NT6000 white-light interferometer and exhibited rms phase variations on the order of $\sigma_\phi\approx9.5$ radians (see discussion below), corresponding to just about the limit of what can be corrected by our DM based on its 3.5$\mu$m peak-to-valley stroke. A characteristic spatial scale of the phase variations was very roughly estimated to be of order $l_\phi\approx1$~mm, leading to an anticipated pupil AO correction range of about $a_0\approx150\mu$m.  

\begin{widetext}

\begin{figure*}[h!]
\centerline{\includegraphics[width=0.9\columnwidth]{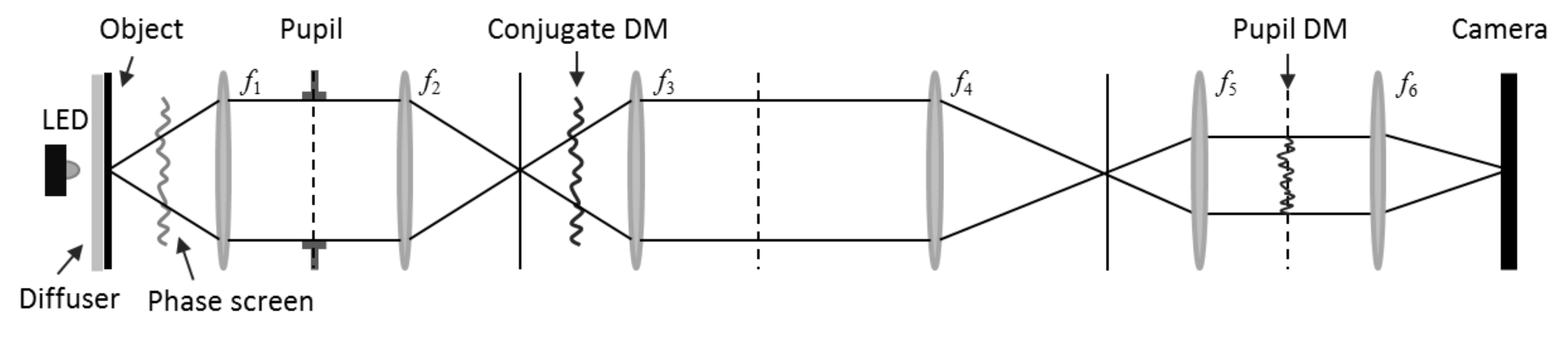}}
\caption{Experimental setup. Lens focal lengths are $f_1=100$~mm, $f_2=100$~mm, $f_3=100$~mm, $f_4=200$~mm, $f_5=75$~mm, and $f_6=50$~mm. Solid vertical lines correspond to object planes; dashed vertical line correspond to pupil planes. Note that in reality the DMs are reflective and the layout was doubly folded. }
\end{figure*}

\end{widetext}

Admittedly, a scenario where a large empty space separates an object from an aberrating layer is unlikely to be found in actual microscopy applications. The purpose of this experiment is only to highlight some salient features underlying conjugate versus pupil AO.

\begin{figure}[h!]
\centerline{\includegraphics[width=.9\columnwidth]{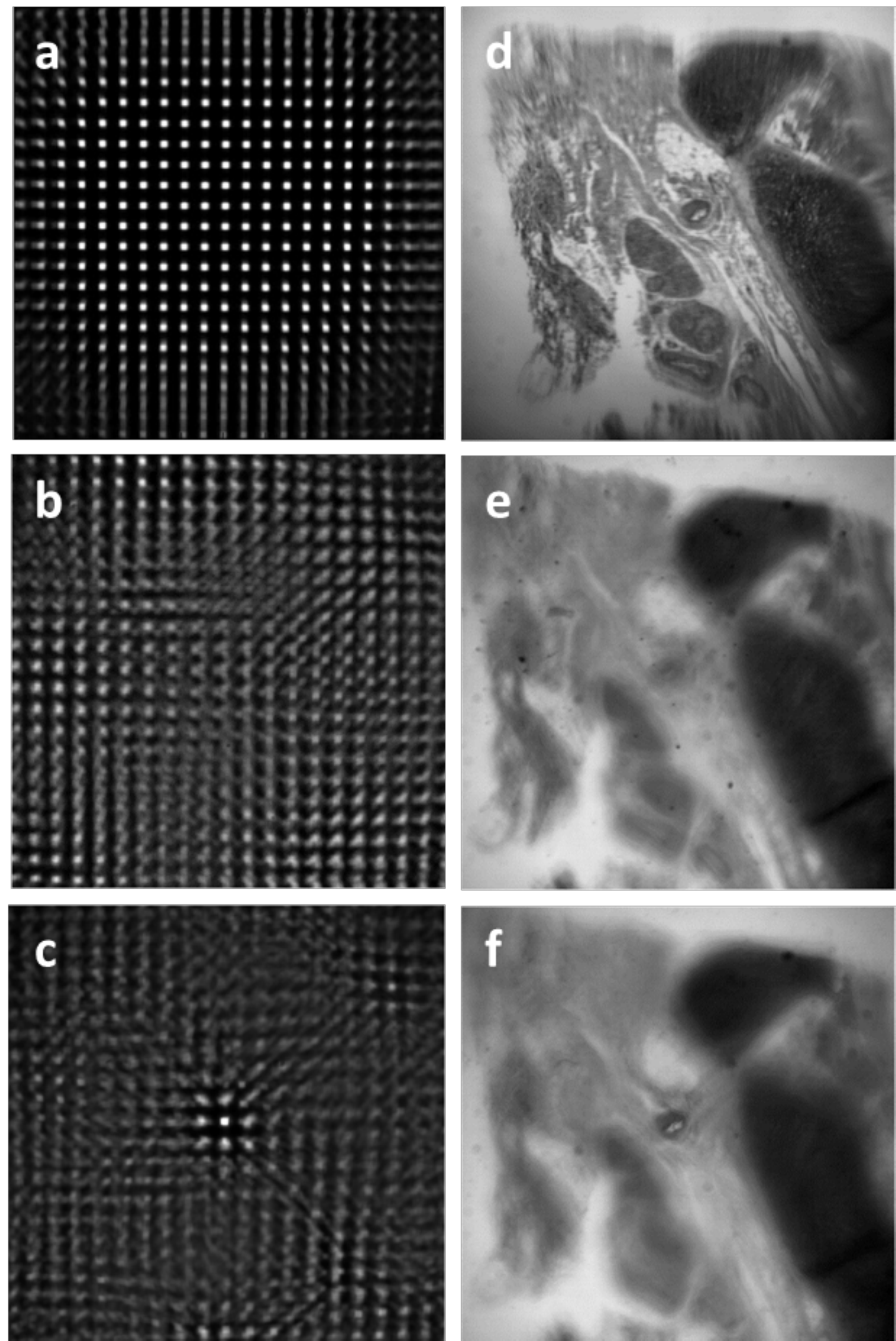}}
\caption{Experimental results for pupil AO. Uncorrected (DM flat) images of an aperture array of period $200\mu$m (a) without and (b) with the presence of an aberrating phase screen.  (c) Same image after AO correction that optimized contrast in a zone about the origin of size $250\times250\mu\textrm{m}^2$ (single guide star). Uncorrected images of tissue section (d) without and (e) with phase-screen, and (f) after AO correction using the same DM pattern as established for (c). Image sizes are $4\times4\textrm{mm}^2$ at the focal plane.       }
\end{figure}

As noted previously, when AO is used in microscopy it is typically applied to laser scanning microscopes where different wavefront corrections are targeted sequentially to different spots distributed throughout the sample. In our case, we wish to apply AO in a non-scanning bright-field configuration. That is, we wish to apply a single wavefront correction to the entire image. To determine an optimized DM shape, we used an image-based stochastic parallel gradient descent (SPGD) procedure that optimizes a particular image metric \cite{vorontsov}. Here, the metric was chosen to be the image contrast measured within a user-selected square zone centered about the image origin (contrast being defined as the square root of the image variance divided by the mean). The application of contrast-based wavefront optimization works well when the object is a localized guide star; however, it is known to fail when the object is arbitrary and extended, since it tends to re-distribute light into mottled patterns rather than improve image sharpness. To circumvent this problem we proceeded in two steps. First, we inserted as an object an array of apertures at the focal plane (see Fig.~7). This calibration object served as a well-defined homogeneous array of guide stars, enabling contrast-based DM optimization to perform adequately. Second, following DM optimization (pupil or conjugate), we replaced the aperture array with an object of interest, namely a thin section of H$\&$E-stained mammal muscle tendon mounted on a microscope slide (Carolina Biological Supply Co.). 

\begin{figure}[h!]
\centerline{\includegraphics[width=.9\columnwidth]{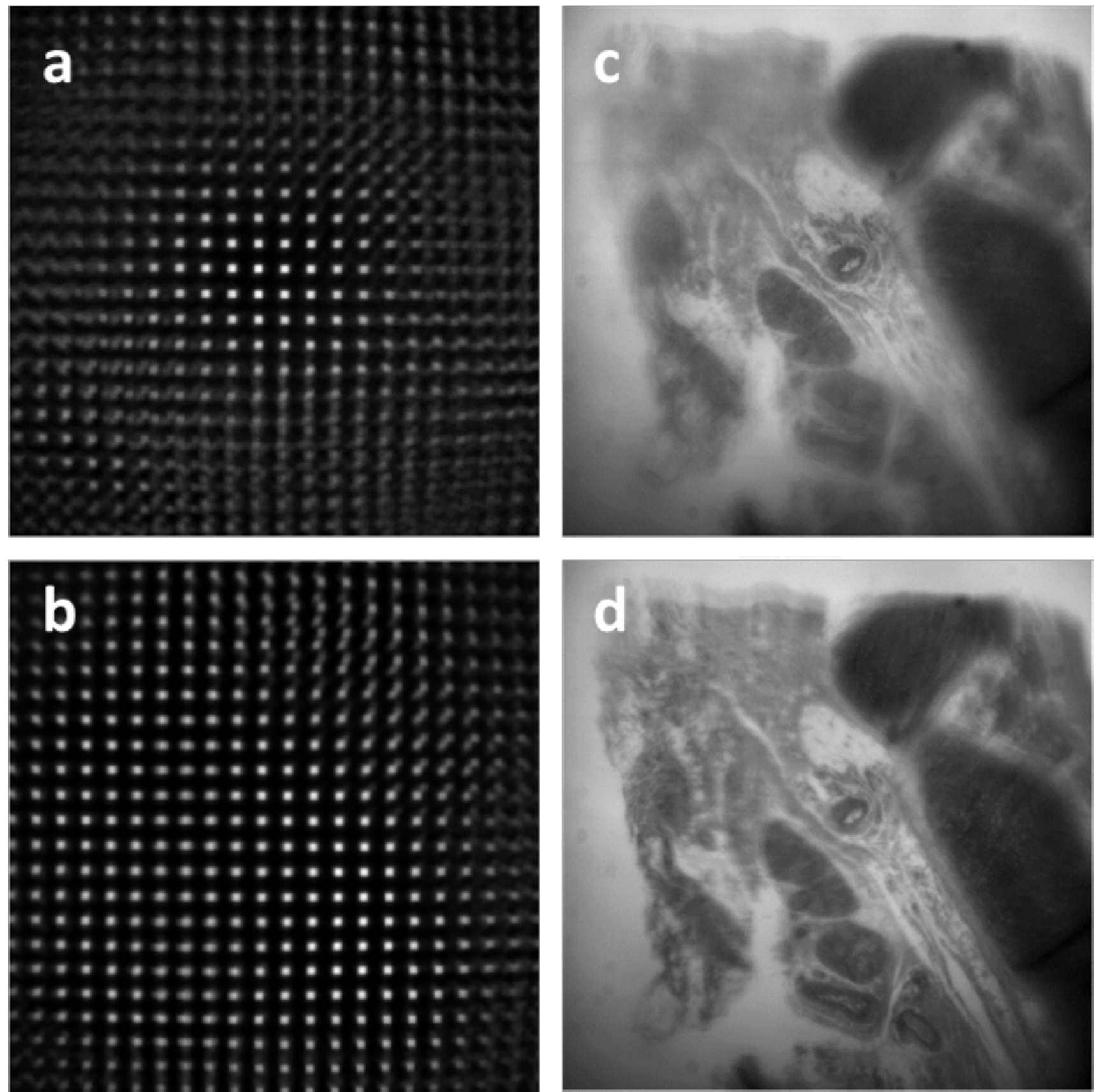}}
\caption{Experimental results for conjugate AO.  Images of aperture array after AO correction that optimized contrast in a zone about the origin of size (a) $250\times250\mu\textrm{m}^2$ (single guide star), and (b) $4\times4\textrm{mm}^2$ (entire image). (c) and (d) are corresponding images of tissue section. }
\end{figure}

\begin{figure}[h!]
\centerline{\includegraphics[width=.9\columnwidth]{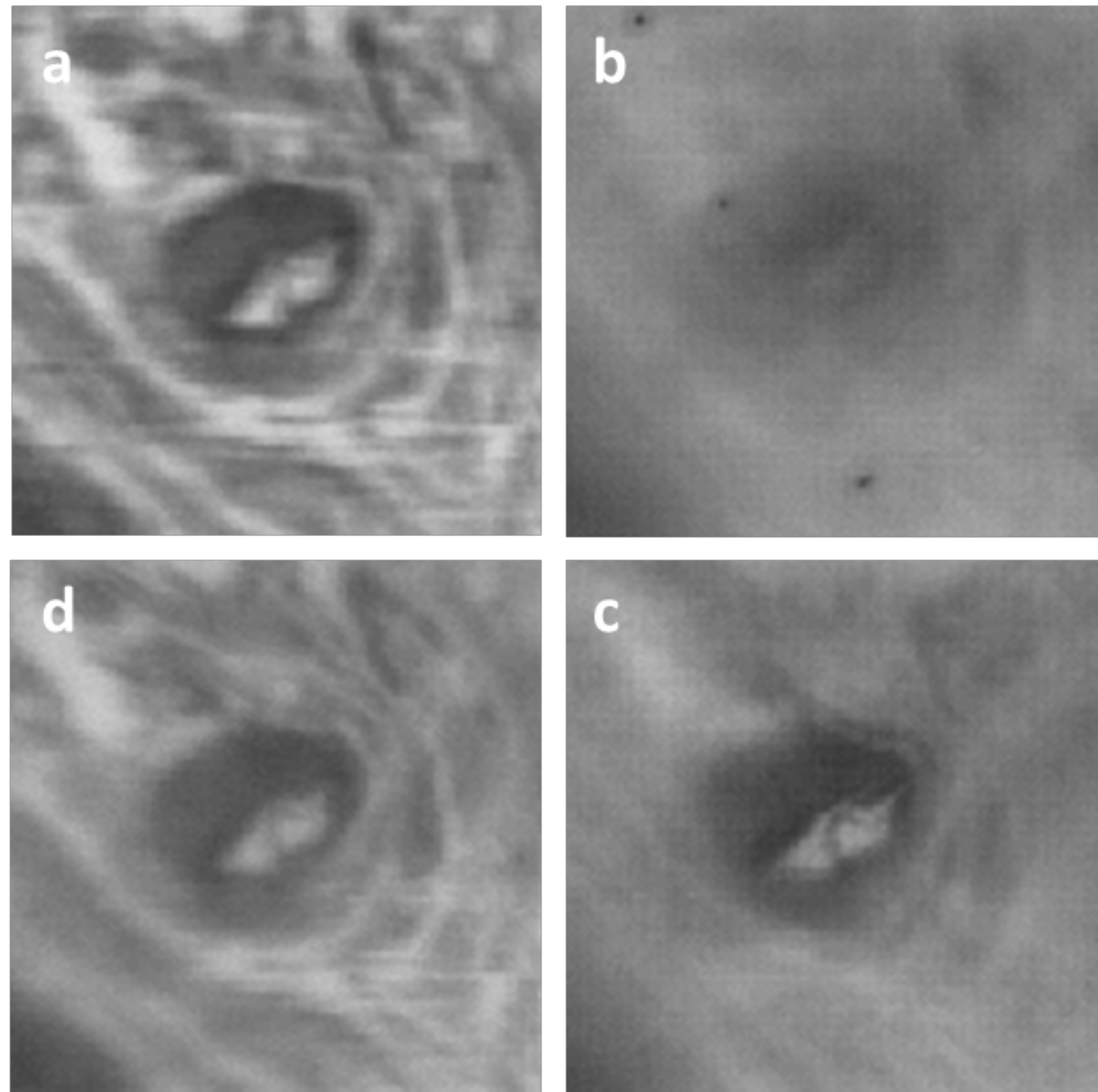}}
\caption{Recapitulation of results. $600\times600\mu\textrm{m}^2$ blowup of zones about the origin taken from images of tissue section (a) uncorrected without aberrations, (b) uncorrected with aberrations, (c) pupil-AO corrected, and (d) conjugate-AO corrected (taken from Fig.~8d).}
\end{figure}

The results for pupil AO are shown in Fig.~7. In this case, the correction zone was chosen to be small, such that it spanned only a single guide star (correction zones larger than this led to DM iterations that failed to converge or increase contrast). Clearly, pupil AO was effective at improving image quality near the origin. However, just as clearly, it was ineffective at improving quality even a small distance from  the origin. The same was true when we inserted the tissue sample, as can be seen in Fig.~7. Based on the properties of the phase screen, the FOV of the wavefront correction was expected  to be about $a_0\approx150\mu$m, which is in rough agreement with experiment .

The results of conjugate AO are shown in Fig.~8. In this case, we chose correction zones that were progressively increased in size (only two of which are shown). As is apparent both with the guide-star array and tissue sample, the resultant FOV increases with correction zone size, approaching the size of the full unaberrated image (Fig.~7a). Of particular interest are Figs. 8a,c. The zone of correction here was $250\times250\mu\textrm{m}^2$, meaning AO was asked to optimize the contrast of the central guide star only. And yet clearly the resultant correction FOV is much larger than this. In this case, the correction FOV corresponds roughly to the region of the phase screen illuminated by the central guide star, as defined by the microscope pupil. That is, it is of size roughly $z$NA. 

\begin{figure}[h!]
\centerline{\includegraphics[width=.8\columnwidth]{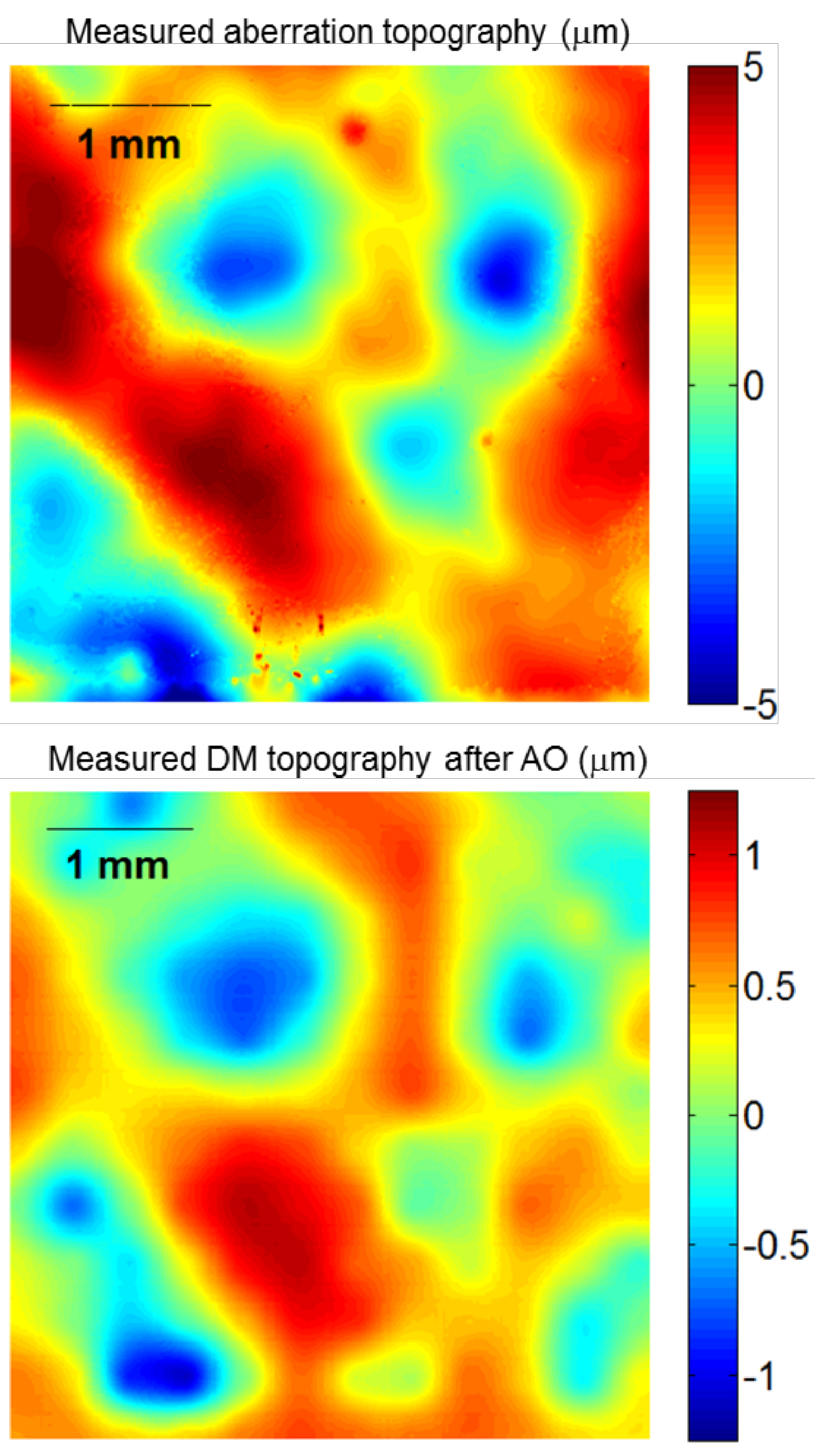}}
\caption{Comparison of the wavefront shapes roughly at the same location of (top) the phase screen and (bottom) the DM after conjugate AO with full-image contrast correction (i.e. same as in Fig.~8b). In both cases, the wavefront shapes were measured with a Zygo white-light interferometer. 
}
\end{figure}

Upon closer inspection of the corrected images near the origin (Fig.~9), we observe that conjugate AO did not provide quite as crisp a correction as pupil AO, One possible reason for this is that only a few conjugate DM actuators contributed to the correction of this small region, whereas for the pupil DM all the actuators contributed. Conjugate AO may have thus suffered from a slight problem of insufficient actuator sampling (a problem that can be corrected with improved sampling). Another possible reason is that edge effects may have undermined the correction efficacy.

Finally, we compare topography maps of the optimized DM shapes with those of the phase screen itself. The correspondence is apparent, both in form and amplitude. Specifically, we note a factor of $\approx4$ difference in topography amplitude between the measured phase screen and DM shapes. This arises in part from the impact of the acrylic topography on its wavefront (the acrylic index of refraction $\approx1.5$ leads to an optical path difference that is about half the local topographic height). Another factor of two arises from the fact that our DM operates in reflection mode, leading to an effective wavefront doubling. In our case the phase-screen aberrations were measured to be 1030~nm rms in wavefront, consistent with an acrylic topography of 2060~nm, and corrected by a DM shape of 514~nm rms. We recall that phase is related to wavefront by $\phi(\brho)=\tfrac{2\pi}{\lambda}W(\brho)$.

\section{Discussion} 
In view of the apparent FOV advantage of conjugate over pupil AO, one may wonder why it is not more prevalent in the microscopy community. There may be several reasons for this. To begin, the most egregious aberrations caused by a sample are often not those induced by laterally spatially variant sample features but rather by a laterally invariant index of refraction mismatch. In this case, conjugate AO does not help and pupil AO is prescribed instead. Moreover, the sample may not be exhibit aberrations in the form of a single, dominant phase screen (as in our idealized, and certainly contrived, demonstration experiment), in which case there may be a difficulty in determining an optimal DM location. Even in the case where there indeed exists a well-defined, dominant aberrating layer, the placement of the DM in its conjugate plane may not be so straightforward. For example, in our experimental demonstration the DM was placed a distance $z$ beyond the nominal microscope image plane (see Fig.~6). However, in the more usual case of a microscope with magnification $M$ much greater than unity, the DM should be placed instead a distance $M^2 z$ from the image plane. For large $M$ this distance may be problematic and require additional re-imaging optics.  There is also the issue of DM actuator size. For pupil AO, the DM actuators should be smaller than $\lambda f/ \zeta$; for conjugate AO they should be smaller than $l_\phi$, which, for weak phase variance, is more restrictive by a factor $\approx z/f$ (magnification notwithstanding). Finally, we must consider how the DM optimization is actually performed.  In our demonstration experiment we had the luxury of being able to swap in a guide-star array to aid iterative image-based optimization. This is generally not possible in practice, and an alternative solution must be found. Ideally, it would be best to measure the aberrations directly using a wavefront sensor; however, this becomes problematic because standard wavefront sensors such as Shack-Hartmann sensors  \cite{tyson} only work well with quasi-collimated wavefronts. To our knowledge, extended-source wavefront sensors (e.g. \cite{monneret, iglesias, PAW}) have not yet been applied to AO in microscopy configurations.

Nevertheless, the above caveats are generally technical in nature. Considering the potentially significant FOV advantage of conjugate AO, its future implementation, perhaps in conjunction with pupil wavefront correction, may well prove to become a new standard in AO applied to microscopy.

\section*{Acknowledgments}
T. Bifano acknowledges a financial interest in Boston Micromachines Corporation. Support for H. Paudel was provided by the Industry/University Cooperative Research Center for Biophotonic Sensors and Systems.

\end{document}